\documentclass[aps,prb,twocolumn,groupedaddress,showpacs]{revtex4}
\usepackage{graphicx}

\begin{document}
\preprint{APS/0000}

\title{Ballistic effects in a proximity induced superconducting diffusive metal}

\author{W. Escoffier}
\email[]{walter.escoffier@manchester.ac.uk}
\thanks{On leave from CEA-DSM-DRFMC-SPSMS, CEA Grenoble, France}
\affiliation{Department of Physics and Astronomy, University of Manchester, Manchester M13 9PL, U.K.}
\author{C. Chapelier}
\email[]{cchapelier@cea.fr}
\affiliation{CEA-DSM-DRFMC-SPSMS, CEA Grenoble, 17 Rue des Martyrs, 38054 Grenoble cedex 9, France}
\author{F. Lefloch}
\affiliation{CEA-DSM-DRFMC-SPSMS, CEA Grenoble, 17 Rue des Martyrs, 38054 Grenoble cedex 9, France}

\date{\today}

\begin{abstract}
Using a Scanning Tunneling Microscope (STM), we investigate the Local Density of States (LDOS) of artificially fabricated normal metal nano-structures in contact with a superconductor. Very low temperature local spectroscopic measurements ($100\ mK$) reveal the presence of well defined subgap peaks at energy $|E| < \Delta$ in the LDOS at various positions of the STM tip. Although no clear correlations between the LDOS and the shape of the samples have emerged, some of the peak features suggest they originate from quasi-particle bound states within the normal metal structures (De Gennes S$^t$James states). Refocusing of electronic trajectories induced by the granular structure of the samples can explain the observation of spatially uncorrelated interference effects in a non-ballistic medium.
\end{abstract}


\maketitle

When a normal metal (N) is in electrical contact with a superconductor (S), it acquires superconducting properties close to its interface: a phenomenon known as the proximity effect. Although this effect was unveiled and well studied in the 1960s within the framework of the Ginsburg-Landau theory \cite{Bogoliubov, DeGennes64}, a revival of interest arose in the 1990s as several groups applied the techniques of mesoscopic physics to fabricate and study hybrid samples on a mesoscopic scale \cite{Gueron, Esteve, Moussy, Vinet}. One of the key concepts relies on Andreev reflection, which provides a way to transfer Cooper pairs from the superconductor into the normal metal or reciprocally electrons from the normal metal into Cooper pairs inside the superconductor \cite{Andreev}. Proximity induced superconducting correlations in N in the vicinity of the S-N interface are of a different quantum nature from those of conventional superconductors. They can be regarded as correlated electron-hole pairs (Andreev pairs) propagating without any attraction potential. In diffusive metals, a microscopic description of the proximity effect emerged, based on the Usadel equation itself derived from the quasi-classical theory \cite{Usadel}. In particular, the Local Density of States (LDOS) of the normal part of the SN junction has been predicted to exhibit a radically different behaviour depending on the characteristic size of N, say, its width $d_n$ in a 1-D model. When the normal metal is semi-infinite ($d_n >> \xi_n$ where $\xi_n=\sqrt{\hbar D_n/\Delta}$ stands for the typical Andreev pairs coherence length, $D_n$ is the electron diffusion constant in N and $\Delta$ the superconducting gap) the LDOS displays a dip near the Fermi energy whose amplitude progressively decreases as the probing distance from the S-N contact increases. However when the normal metal is finite ($d_n < \xi_n$), a position independent mini-gap appears, whose width is of the order of the Thouless energy defined as $E_{Th}=\hbar D_n / d_n^2$ \cite{Belzig, Gupta}. On the other hand, the properties of S-N bilayers are predicted to change radically from the above description if the normal metal is a ballistic medium. Andreev's pairs propagating straight and bouncing back and forth in the normal part of the junction give rise to quantum bound states (coined De-Gennes S$^t$James states \cite{DeGennes63}), which yield peaks at energies $E_k < | \Delta |$ in the density of states of the system. Clearly, such peaks result from interference effects which, to our knowledge, have never been studied with a thermal resolution much lower than $\Delta$. The first observation of De-Gennes S$^t$James resonances have been reported by Tessmer {\it et. al.} \cite{Tessmer} who measured small sub-gap oscillations in the local density of states of NbSe$_2$/Au contacts with the help of a Scanning Tunneling Microscope (STM) at $1.6\ K$. Two years later, Levi {\it et. al.} \cite{Levi} observed similar effects in diffusive NbTi/Cu junctions at $4.2\ K$. Even if the temperature-limited resolution of their experiment restricted its extent, it appears that ballistic effects still can be seen in diffusive metals.

In this paper, we investigate the LDOS characteristics of artificially fabricated diffusive SNS systems of mesoscopic dimensions using a very low temperature STM. We observed sub-gap peaks of a similar nature to those reported in Levi {\it et al}, but energetically better resolved since the temperature was only $100\ mK$. These peaks are believed to be related to quantum bound states involving the quasi-ballistic motion of electrons in N.\\

\begin{figure}
\includegraphics[width=7.0cm]{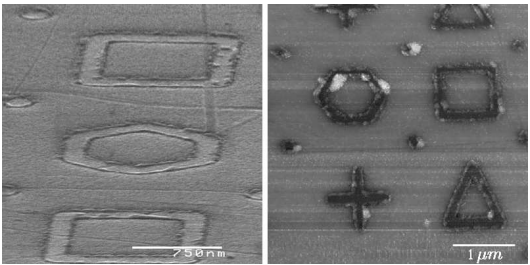}
\caption{\label{figure1} Room temperature images of parts of the sample: a) tilted SEM image and b) STM image. Here the niobium layer has been slightly over-deposited and the overall sample surface is not perfectly flat. Some electro-resist bits could not be removed and appear as white "bubbles" in STM images. They sometimes can alter the quality of the STM images at low temperature, preventing an accurate location of the SN interface.}
\end{figure}

Samples consist of mesoscopic structures of normal metal (Au) embedded by a superconducting matrix (Nb). They were designed as dots, crosses, squares, triangles or hexagons with a side length of about $1\ \mu m$. Their width and thickness are $200\ nm$ and $20\ nm$ respectively (see Fig. \ref{figure1} or Fig. \ref{figure4} for a schematic representation of a square-shaped structure). These structures are placed in the center of a $0.5 \times 0.5\ mm^2$ lithographed test card to allow their quick location using the STM X-Y displacement stage, prior to cooling down the system. Samples were fabricated using electron beam lithography (with negative photo-resist) on an initial $20\ nm$ sputtered gold layer on $Si$. Argon etching, through the lithographed photo-resist mask, has been used to create the structures, before sputter-depositing a $20\ nm$ thick niobium film. Mechanically assisted lift-off of niobium on top of the gold structures and cleaning are the final steps of the process. This fabrication method has been developed in order to fulfill several requirements:\\
(i) The interface between the superconductor and the normal metal is transparent enough with no significant barrier in order to yield an Andreev reflection coefficient close to 1 and a fully developed superconducting proximity effect \cite{Blonder}.\\
(ii) Niobium superconducting properties at the free surface of the samples are not damaged by the fabrication process and a gap in the LDOS close to its bulk value is measured on top of the film \cite{Hoss}.\\
(iii) Sample roughness and cleanliness is compatible with the STM imaging capabilities.

Although much care is taken during cleaning, some photo-resist residues remain difficult to remove. We systematically tried to avoid the dirtiest regions of the sample when positioning the STM tip and since no peculiar effects have ever been observed when tunneling close to the residues, we believe they do not affect our conclusions. The electronic mean free path of gold, extracted from resistivity measurements of films made in the same conditions as mentioned earlier, has been measured to be approximately $22\ nm$. Assuming a Fermi velocity $v_F=1.39\times10^6\ m.s^{-1}$, the diffusion constant $D_n$ is about $100\ cm^2.s^{-1}$. These films show a granular structure with a mean grain diameter of $50\ nm$.\\

\begin{figure}
\includegraphics[height=3.5cm]{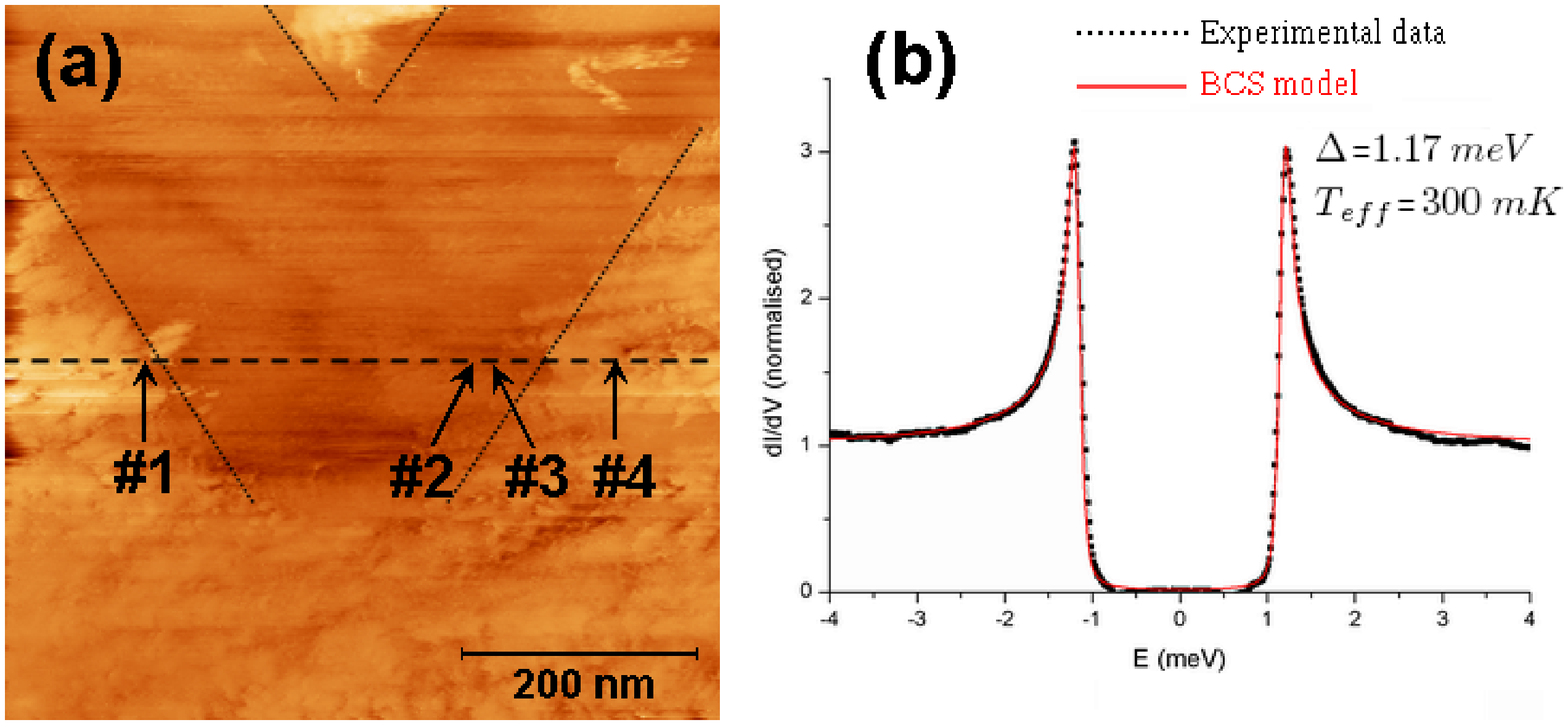}
\includegraphics[width=8cm]{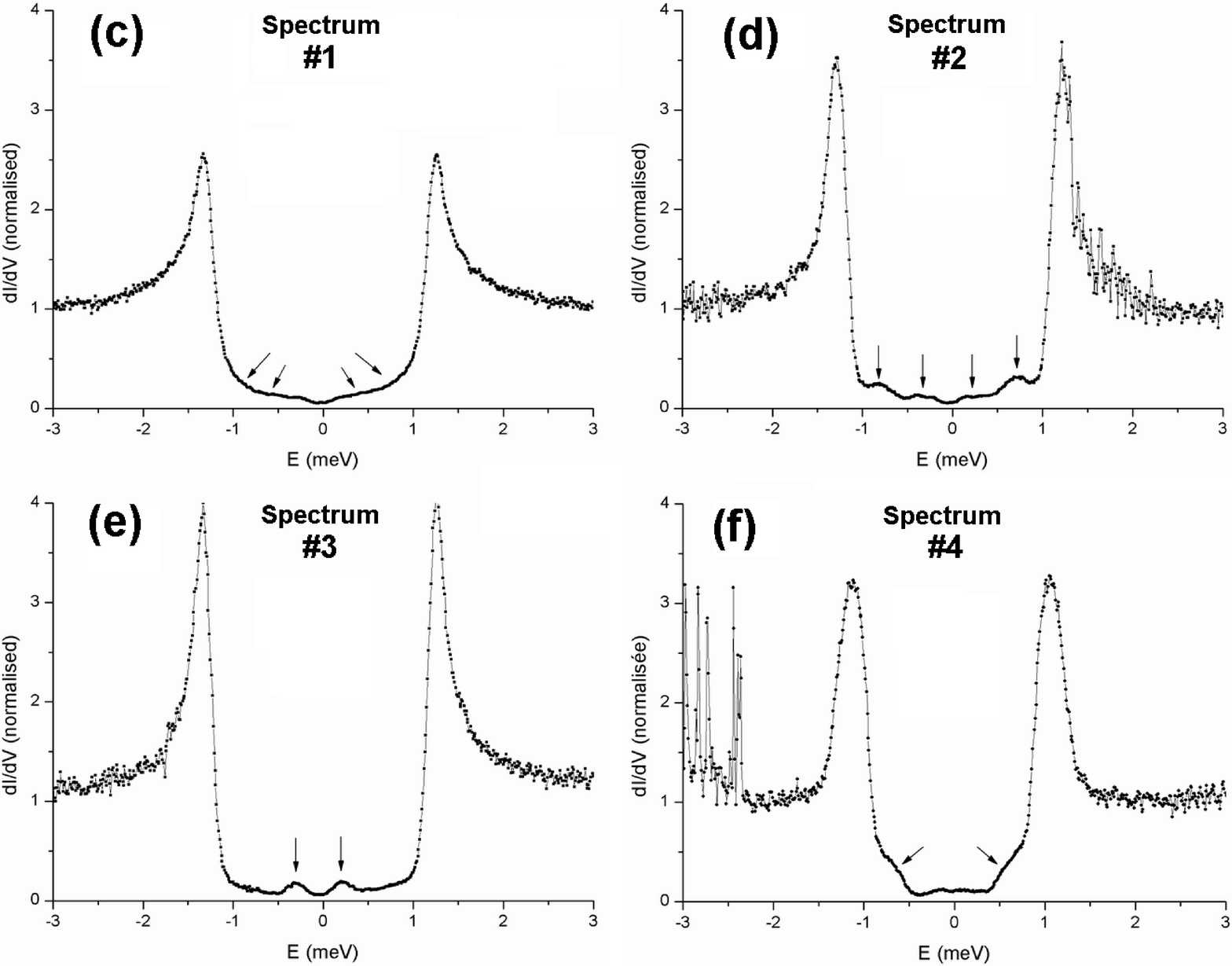}
\caption{\label{figure2} Spectra have been recorded automatically along the horizontal dashed line in the center of the image, crossing one end of a triangle-like structure whose boundaries have been marked by the dotted lines. Only the most significant spectra are displayed, at the positions indicated by the arrows.}
\end{figure}

We used a home-made STM mounted in a dilution fridge, operating at a base temperature of $100\ mK$, to measure the local density of states of the normal metal structures at various positions. A small AC modulation of $20\ \mu V$ rms was added to the sample-tip DC bias voltage $V$ and the differential conductance $\frac{dI}{dV}(V)$ was obtained with a lock-in amplifier technique. Depending on experimental parameters, a typical spectrum acquisition lasts between 5 and 10 s. Noisy or unphysical spectra are automatically removed from analysis. All spectra are renormalised so that the flat density of states at energies far from $\pm \Delta$ equals 1 . When tunneling above niobium, spectra can be reproduced well using the BCS model for the density of states (see Fig. \ref{figure2}(b)), with $\Delta=1.17\ meV$ and an effective temperature of $T_{eff}=300\ mK$ (suggesting an electron heating effect by unfiltered electromagnetic radiation).

Now we would like to focus on spectra acquired when tunneling above the normal metal structures. A STM topographic image of one of the vertices of a triangle-like structure is shown in Fig. \ref{figure2}(a). A series of spectra has been acquired on 256 points spaced by $2.3\ nm$ along the horizontal dashed line. Four consecutive spectra have been automatically acquired on each spot in order to check their reproducibility. Near the middle of the structure, most of the spectra show weakened BCS-like LDOS similar to the typical niobium spectra reported above, indicating a strong proximity effect. A few of them however show small symmetric peaks with respect to the Fermi energy $E_F=0$ inside the gap (e.g. spectra of Fig. \ref{figure2}(d) and (e)). Although these peaks remain almost identical in the four spectra acquired consecutively at the same position, their number, energetic position and amplitude can change radically when the STM tip is slightly moved (e.g. spectra of Fig. \ref{figure2}(d) and (e) have been recorded only $17\ nm$ apart but show very different peak configurations). As discussed later, we suggest these peaks originate from interfering Andreev's pairs trajectories in the normal metal. Isotropic diffusion or inelastic scattering effects affect the Andreev's pairs coherence and explain the small amplitude of the peaks. Next to the SN boundaries, we observed non-BCS spectra, similar to spectra of Fig. \ref{figure2}(c) and (f), having a non-zero density of states at energies inferior, but close to $\pm \Delta$. We have no definitive interpretation for the general shape of these spectra yet. However, we suggest that such states in the LDOS involve numerous, short and poorly defined different electronic trajectories connected to the SN interface which yield a broad distribution of the aforementioned peaks below the gap edges.
\\
In a separate experiment, we investigated the LDOS of square-shaped structures. Although STM images were rather clear at the beginning of the experiment, their quality kept decreasing on cooling down the system, preventing us from accurately correlating spectroscopic and topographic information. Nevertheless, it did not seem to affect the quality of spectroscopic measurements, which at different locations of the STM tip show alternately BCS-like density of states and spectra with very well-defined sub-gap peaks, similar to, but sharper and higher than those reported earlier in the triangle structure. On the other hand, peaks turned out to be very stealthy, as several consecutive spectral acquisitions at the same nominal position failed to display the same peak configuration (energetic position, amplitude and number). We interpret this effect as an extreme sensitivity of the peaks to the tunneling position, as discussed later. The relatively large number of spectra showing such peaks (75 selected spectra in total) allows us to summarize their characteristic features:\\
(i) Only even numbers of peaks (up to six) have been observed.\\ 
(ii) Energetic peak positions are always symmetric with respect to the Fermi energy and to a lesser extent in amplitude.\\
(iii) Apart from a couple of spectra which show peaks at energies $\pm E_k$ slightly above $\Delta$, the majority of them lie within the energy gap.

To our knowledge such strong peaks have never been observed previously and we never saw them in our samples when we performed similar experiments above 1K. It is the reason why we believe they result from electronic interference effects, where long coherence lengths are required.\\

Although peaks appear to be position-dependent, the observation of quasi-similar peaked spectra on both the triangle-shaped and square-shaped structures indicates that the macroscopic geometry has very little influence on the LDOS. Nevertheless, the above mentioned characteristics of the peaks support the hypothesis they result from resonant De Gennes S$^t$ James states due to the confinement of Andreev pairs inside the normal metal structure.
\begin{figure}
\includegraphics[width=8cm]{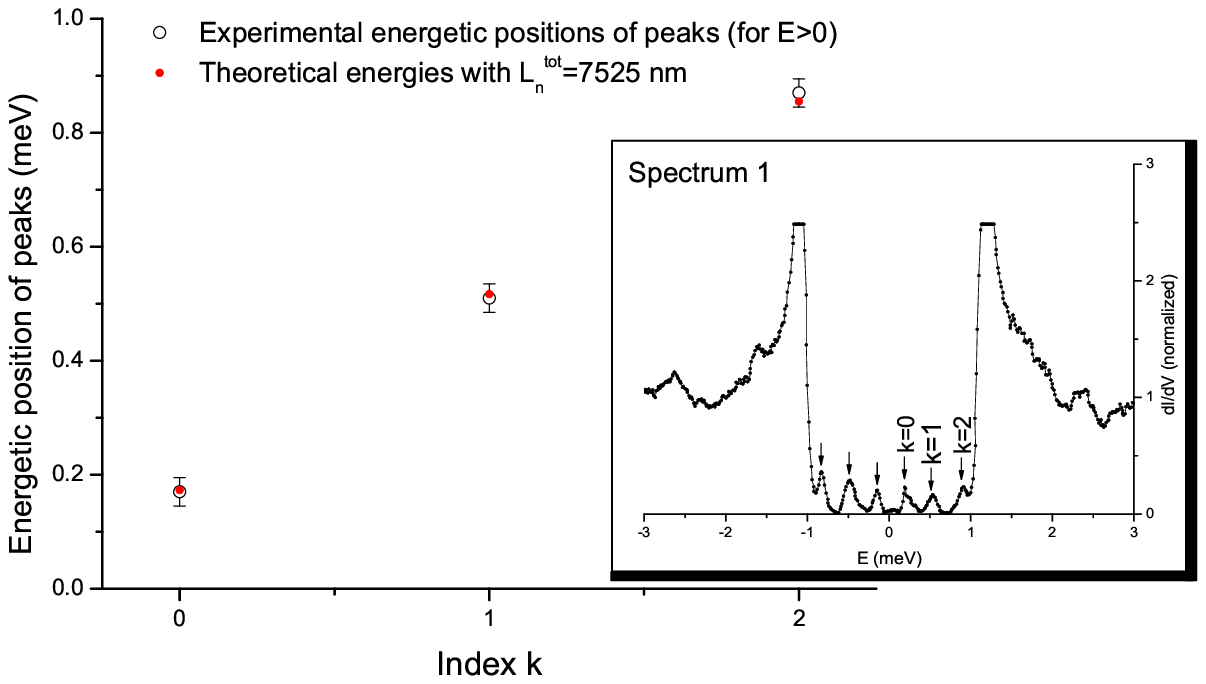}
\includegraphics[width=8cm]{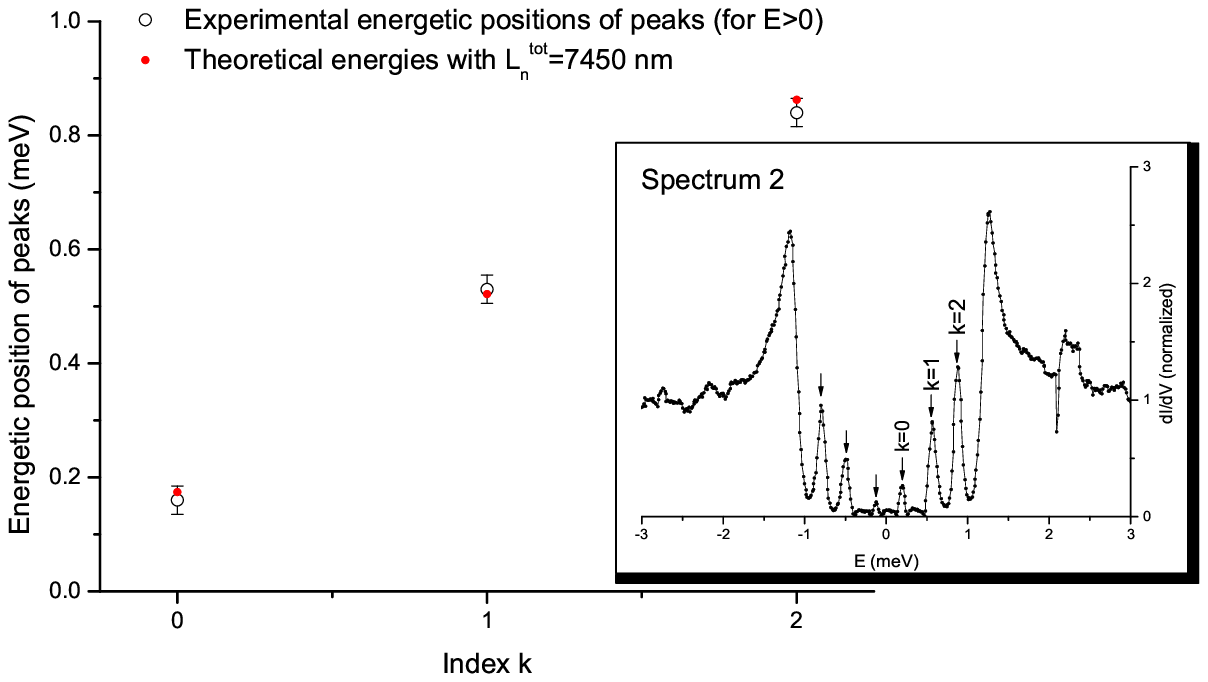}
\caption{\label{figure3} Peak energetic positions reported for two typical spectra (shown in inserts) acquired on one side of the square-shaped normal metal structures. For $E > 0$, three peaks are clearly visible in both spectra and have been labeled with index $n=0,1,2$. Solutions of the De Gennes S$^t$James equation have been displayed with adjusted trajectory lengths $L^{tot}_n=7525\ nm$ for spectrum 1 and $L^{tot}_n=7450\ nm$ for spectrum 2.}
\end{figure}
Spectra containing a large number of peaks offer the opportunity to study their relative energetic positions. Fig. \ref{figure3} shows two spectra with six peaks each. Restricting our analysis to positive energies (peaks are symmetric), their labeled energetic positions $E^{exp}_k$ (n=0,1,2) have been reported in table \ref{table1}.

On the other hand, the De Gennes $S^t$ James quantification equation in the case of a SNS system states:
\begin{equation}
\frac{E_k L^{tot}_n}{\hbar v_F}=k \pi + arcos\left(\frac{E_k}{\Delta}\right)
\end{equation}
where $L^{tot}_n$ corresponds to the length of the interfering Andreev pairs trajectory. Using $\Delta=1.17\ meV$ and $v_F=1.39\times 10^6\ m.s^{-1}$, $L^{tot}_n$ is the only adjustable parameter that has to be tuned in order to match the set of $E_k$ with the experimental ones. A good agreement has been achieved using $L_n^{tot}=7525\ nm$ for spectrum 1 and $L_n^{tot}=7450\ nm$ for spectrum 2 (see table \ref{table1}).
\begin{table}
	\centering
		\begin{tabular}{|c||c||c|}
		\hline
     & Spectrum 1 & Spectrum 2 \\
    \hline
		\begin{tabular}{c} Index\\0\\1\\2 \end{tabular} & \begin{tabular}{c|c} $E_k$ & $E_k^{exp}$\\\hline$0.17$ & $0.17$\\$0.52$ & $0.51$ \\ $0.85$ & $0.87$ \end{tabular} & \begin{tabular}{c|c} $E_k$ & $E_k^{exp}$\\\hline$0.17$ & $0.16$\\$0.52$ & $0.53$\\$0.86$ & $0.84$ \end{tabular}\\
		\hline
		\end{tabular}
\caption{\label{table1} Theoretical and experimental energies (meV) of the peaks for spectrum 1 and 2 of Fig. \ref{figure3}.}
\end{table}
This striking agreement raises the question of the nature of the electron-hole trajectories in the diffusive gold structure. Since the mean free path and the grain diameter are of the same order of magnitude, it is very likely that most of the scattering events occur predominantly at grain boundaries. Because these extended defects are much bigger than the Fermi wavelength, we expect specular reflection or very anisotropic diffusion to dominate at least for specific trajectories. Therefore, the absence of isotropic quantum diffraction on point-like impurities allows us to consider quasi-ballistic paths in a semi-classical picture as represented in Fig. \ref{figure4}. The very existence of such a trajectory bouncing back and forth between SN boundaries with an unfolded total length $L^{tot}_n$ can explain why the observed experimental peaks are well described by the ballistic De Gennes S$^t$ James model. Moreover, the corresponding diffusion length $d^{diff}_n=\sqrt{D_nL^{tot}_n/v_F}$ in N is of the order of $230\ nm$, which is close to the nominal width $d_n=200\ nm$ of the normal metal structures.
\begin{figure}
\includegraphics[width=7cm]{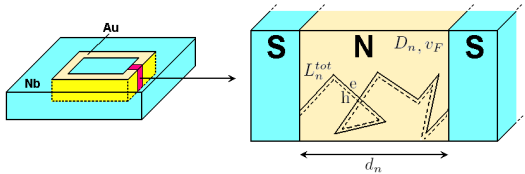}
\caption{\label{figure4} Schematic representation of the system under study. $d_n=200\ nm$ is the nominal width of the sides of the square-shaped normal metal structure while $L^{tot}_n$ corresponds to the total length of the Andreev pair trajectory connecting the two SN interfaces.}
\end{figure}
The above analysis considers a unique Andreev pair and its associated trajectory, satisfying the conditions for an interference effect to occur. According to reference \cite{DeGennes63}, the presence of many trajectories of different length results in a peak broadening effect, associated with a non-zero sub-gap density of states. This theoretical work however considers the spatially integrated density of states of a SN slab. In our case, the STM tip constitutes a local probe and implies a selection among all the possible Andreev pairs' trajectories. This explains why we observed very well defined peaks, instead of the sawtooth structure of the density of states initially proposed in reference \cite{DeGennes63}. Furthermore, according to reference \cite{Shytov}, the discreteness of the peaks can also result from a localization effect of the Andreev pair induced by the granular structure of the samples: grains could act as lenses which re-focus diverging hole and electron trajectories of the Andreev pair and enhance interference effects. If the STM tip is in the vicinity of a localized state, it will give rise to a peak in the LDOS. As grains are randomly distributed, authors of ref. \cite{Shytov} predict that peaks measured by STM should be spatially uncorrelated, as indeed we observed. To a further extent, it could also explain the reported high sensitivity of the peaks' configuration between two successive spectroscopic acquisitions at the same nominal position. Actually, between these two acquisitions, the feedback loop of the STM is activated for a short time: this can result in a very small uncontrolled shift of the STM tip position and drive the system out of resonance.\\

To conclude, we report the observation of sharp sub-gap peaks in the local density of states of S-N systems at $100\ mK$ using the scanning tunneling spectroscopy technique. Their characteristics can accurately be described by De Gennes S$^t$James resonant states in the normal metal islands, although the short electronic mean free path of the samples should normally prevent their existence. If these peaks are indeed due to electronic interference effects, this would bring about new fundamental questions concerning the limit between ballistic, quasi-ballistic and diffusive systems.

\end{document}